# Adaptive Non-Rigid Inpainting of 3D Point Cloud Geometry

Chinthaka Dinesh, *Student Member, IEEE,* Ivan V. Bajić, *Senior Member, IEEE,* and Gene Cheung, *Senior Member, IEEE*

*Abstract*—In this letter, we introduce several algorithms for geometry inpainting of 3D point clouds with large holes. The algorithms are examplar-based: hole filling is performed iteratively using templates near the hole boundary to find the best matching regions elsewhere in the cloud, from where existing points are transferred to the hole. We propose two improvements over the previous work on exemplar-based hole filling. The first one is adaptive template size selection in each iteration, which simultaneously leads to higher accuracy and lower execution time. The second improvement is a non-rigid transformation to better align the candidate set of points with the template before the point transfer, which leads to even higher accuracy. We demonstrate the algorithms' ability to fill holes that are difficult or impossible to fill by existing methods.

## I. Introduction

A 3D point cloud consists of a set of points in 3D space, sometimes with attributes such as color. Point clouds are used for representing the geometry of 3D objects and scenes. Recent advances in 3D scanning technologies are making 3D point clouds popular in augmented reality, mobile mapping, gaming, 3D telepresence, scanning of historical artifacts, and 3D printing. However, the scanned 3D point cloud may be missing data in certain regions due to occlusion, low reflectance of the scanned surface, limited number of scans from different viewing directions, etc. [1]. In certain applications such as remote telepresence, parts of the point cloud may be lost due to unreliable communication links along the way. Hence, filling in the missing regions (holes) is an important problem in 3D point cloud processing. There are a number of methods for detecting holes in point clouds based on variations of the point density [2], [3], [4]. In remote telepresence, point cloud data is ordered and packetized for transmission to a remote location. Here, missing packets' indices can be used to determine the locations in 3D space where the missing data used to be. In this letter, we assume that the hole has been identified using one of these existing methods.

Various methods have been proposed for hole filling of surfaces represented by meshes [5], [6], [7], [8], but the related problem of hole filling for 3D point clouds has received less attention. In [9], a triangle mesh is constructed from the input point cloud, then the missing region is interpolated using a moving least squares approach. However, this method produces unsatisfactory results on large holes, especially if the underlying surface is complex. In [10], a plane tangent to the missing part is determined and the hole boundary is projected onto this plane. Then, its convex hull is computed and points are created in such a way that the sampling allows to cover a dilated version of the convex hull. Next, a $k$-nearest neighbors graph is constructed from the union of input point cloud and this set of created points. Finally, a graph-based Partial Differential Equation (PDE) solver is used to deform the generated points to fit into the hole. This method gives good results for small holes or smooth surfaces, but it faces problems if the hole boundary contains folds or twist. In [2], a hole-filling method based on the tangent plane for each hole boundary point is proposed. Traversing the boundary in a clock-wise direction, points on each tangent plane are computed and inserted into the hole. This process is repeated from the hole boundary towards the interior. While this works for small holes, it faces the same problems as [9], [10] on large holes and complex surfaces.

A few methods have been proposed to fill relatively large holes in 3D point clouds. In [11], a non-iterative framework based on Tensor Voting is proposed to fill-in holes by using neighbourhood surface geometry and geometric prior derived from registered similar models. In [12], a hole is filled by propagating local 3D surface smoothness from around the hole by harvesting the cues provided by a similar model. Although they can fill large holes reasonably well, both methods need at least one complete model similar to the model with the hole. Recently, we proposed an exemplar-based framework [13] for hole filling, which exploits non-local self similarity to provide plausible reconstruction even for large holes and complex surfaces. This approach is inspired by [14] and its success in image inpainting. The main focus of [13] was on the technical challenges involved in transplanting the image-based method from [14] into the point-cloud framework.

In this letter, two novel concepts are proposed to improve our previous framework [13]. The first one is adaptive template size selection to match the local surface characteristics. This leads to both higher accuracy and lower execution time compared to [13]. The second improvement is a non-rigid transformation to better align the set of points that will be transferred into the hole with the surface characteristics near the hole. This improvement further increases the accuracy.

A brief review of our previous hole-filling method [13] is presented in Section II, followed by the proposed improvements in Section III, experimental comparison with [13], [9],

C. Dinesh and I. V. Bajić are with the School of Engineering Science, Simon Fraser University, Burnaby, BC, Canada, e-mail: hchintha@sfu.ca and ibajic@ensc.sfu.ca.

G. Cheung is with National Institute of Informatics, Tokyo, Japan, e-mail: cheung@nii.ac.jp.

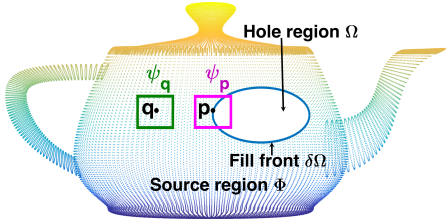

Fig. 1. Illustration of the source region $\Phi$, hole $\Omega$ and fill front $\delta\Omega$.

[10], [2] in Section IV, and conclusions in Section V.

## II. BRIEF REVIEW OF [13]

The set of available points in the point cloud is called the *source region*, denoted $\Phi$. The hole is denoted $\Omega$, and its boundary is denoted $\delta\Omega$. The boundary (*fill front*) evolves inwards as hole filling progresses. A cube centered at point $\mathbf{q}$, with edges parallel to the $x, y, z$ axes, is denoted $\psi_\mathbf{q}$. Fig. 1 illustrates these concepts. The size of each cube in [13] is $10 \times 10 \times 10$ voxels. Here, voxel is a cubic unit of space. To determine the voxel size for a given input cloud, we perform octree partitioning until each cloud point resides in its own voxel, then the smallest voxel size in the octree is taken as the voxel size for that cloud. Octree partitioning is started with a cubic bounding box that encapsulates the whole point cloud. If the partitioning stops after $n$ steps, the volume of the smallest voxel is $2^{-3n}$ of the volume of the original bounding box. Hole filling consists of three steps – priority calculation, template matching, and point transfer – applied iteratively until the hole is filled. Each step is explained below.

**Priority Calculation:** The first step is to assign priority to each $\psi_\mathbf{p}$ for $\mathbf{p} \in \delta\Omega$. The priority is biased towards those $\psi_\mathbf{p}$'s that seem to be on the continuation of ridges, valleys, and other surface elements that could more reliably be extended into the hole. For a given $\psi_\mathbf{p}$, the priority $P(\mathbf{p})$ is defined as

$$P(\mathbf{p}) = D(\mathbf{p})C(\mathbf{p}), \qquad (1)$$

where $D(\mathbf{p})$ is the data term, which depends on the structure of the data in $\psi_\mathbf{p}$ as discussed in [13], and $C(\mathbf{p})$ is the number of available points in $\psi_\mathbf{p}$ ($C(\mathbf{p}) = |\psi_\mathbf{p} \cap \Phi|$).

**Template Matching:** Once all priorities on the fill front have been computed, the highest priority cube $\psi_{\widehat{\mathbf{p}}}$ is selected as $\widehat{\mathbf{p}} = \arg\max_{\mathbf{p} \in \delta\Omega} P(\mathbf{p})$. The available points in $\psi_{\widehat{\mathbf{p}}}$ are called the *template*. Then, the source region is searched for the cube $\psi_\mathbf{q}$ ($\psi_\mathbf{q} \subset \Phi$) that best matches this template. As discussed in [13], many candidate cubes can be eliminated during the search to speed up the process. To find the best match for $\psi_{\widehat{\mathbf{p}}}$ among the candidate cubes, $\psi_\mathbf{q}$ is first translated such that point $\mathbf{q}$ coincides with $\widehat{\mathbf{p}}$. This translated $\psi_\mathbf{q}$ is denoted $\overline{\psi}_\mathbf{q}$. Then, the best 3D rotation matrix $\mathbf{R}_b$ is determined to align $\overline{\psi}_\mathbf{q}$ with $\psi_{\widehat{\mathbf{p}}}$, and the rotated cube is denoted $\overline{\psi}_\mathbf{q}^{\mathbf{R}_b}$. Specifically, the rotation matrix is found as

$$\mathbf{R}_b = \arg\min_{\mathbf{R}} d\left(\overline{\psi}_\mathbf{q}^{\mathbf{R}}, \psi_{\widehat{\mathbf{p}}}\right), \qquad (2)$$

where $\mathbf{R}$ is a 3D rotation matrix and $d(\overline{\psi}_\mathbf{q}^{\mathbf{R}}, \psi_{\widehat{\mathbf{p}}})$ is the One-sided Hausdorff Distance (OHD) [15] from $\psi_{\widehat{\mathbf{p}}}$ to $\overline{\psi}_\mathbf{q}^{\mathbf{R}}$,

$$d\left(\overline{\psi}_\mathbf{q}^{\mathbf{R}}, \psi_{\widehat{\mathbf{p}}}\right) = \max_{\mathbf{a} \in \psi_{\widehat{\mathbf{p}}}} \min_{\mathbf{b} \in \overline{\psi}_\mathbf{q}^{\mathbf{R}}} \|\mathbf{a} - \mathbf{b}\|_2, \qquad (3)$$

where $\|\cdot\|_2$ is the Euclidean norm.

The Iterative Closest Point (ICP) algorithm [16] is used to find $\mathbf{R}_b$ in (2) for each candidate cube $\overline{\psi}_\mathbf{q}$, and the corresponding OHD after alignment, $d(\overline{\psi}_\mathbf{q}^{\mathbf{R}_b}, \psi_{\widehat{\mathbf{p}}})$, is recorded. According to [16], the complexity of finding $\mathbf{R}_b$ is $\mathcal{O}(N_1 \log N_2)$, where $N_1$ and $N_2$ are the numbers of points in $\psi_{\widehat{\mathbf{p}}}$ and $\overline{\psi}_\mathbf{q}^{\mathbf{R}_b}$, respectively. Then the rotated cube with the smallest aligned OHD, $\overline{\psi}_{\widehat{\mathbf{q}}}^{\mathbf{R}_b}$, is selected for transfer. Specifically,

$$\overline{\psi}_{\widehat{\mathbf{q}}}^{\mathbf{R}_b} = \arg\min_{\overline{\psi}_\mathbf{q}^{\mathbf{R}_b}} d\left(\overline{\psi}_\mathbf{q}^{\mathbf{R}_b}, \psi_{\widehat{\mathbf{p}}}\right). \qquad (4)$$

**Point Transfer:** First, the points of $\overline{\psi}_{\widehat{\mathbf{q}}}^{\mathbf{R}_b}$ are matched with those of $\psi_{\widehat{\mathbf{p}}}$. Let $\mathbf{x}_i \in \psi_{\widehat{\mathbf{p}}}$ be a point in $\psi_{\widehat{\mathbf{p}}}$. Let $\mathbf{y}_i$ be the closest (by Euclidean distance) point to $\mathbf{x}_i$ within $\overline{\psi}_{\widehat{\mathbf{q}}}^{\mathbf{R}_b}$,

$$\mathbf{y}_i = \arg\min_{\mathbf{y} \in \overline{\psi}_{\widehat{\mathbf{q}}}^{\mathbf{R}_b}} \|\mathbf{y} - \mathbf{x}_i\|_2. \qquad (5)$$

Then we say that $\mathbf{y}_i$ has been *matched* with $\mathbf{x}_i$ and we add it to the set of matched points of $\overline{\psi}_{\widehat{\mathbf{q}}}^{\mathbf{R}_b}$, denoted $M_{\widehat{\mathbf{q}}}$. Finally, all the unmatched points, $\overline{\psi}_{\widehat{\mathbf{q}}}^{\mathbf{R}_b} \setminus M_{\widehat{\mathbf{q}}}$ are transferred to $\psi_{\widehat{\mathbf{p}}}$ and the fill front is updated. This completes one iteration of the filling procedure; after this, the fill front $\delta\Omega$ is updated and the process repeated until the hole is filled. The filling stops at the iteration in which the template covers the entire fill front.

## III. PROPOSED METHODS

Although [13] outperforms existing methods such as [9], [10], [2], there is still room for improvement. Two such improvements are described here: 1) adaptive cube size (ACS) for template selection, and 2) non-rigid transformation (NRT) for improved alignment between the template and the best-matched cube. With those modifications, the size of the template cube $\psi_{\widehat{\mathbf{p}}}$ will change depending on the surface characteristics, and a non-rigid transformation will be applied to the best-matched candidate cube prior to point transfer. Details are provided in the following sections.

### A. Adaptive Cube Size

The fixed template cube size, as used in [13], may be inappropriate in several cases. For example, when the surface is relatively flat near $\widehat{\mathbf{p}}$, small cube size may lead to finding an inaccurate match elsewhere in the model where the surface is also relatively flat. In such a case, the template needs to be increased in order to capture sufficiently discriminative surface structure. There are also other cases where the surface characteristics near $\widehat{\mathbf{p}}$ are so complicated that even the best-matched cube of a given size is poorly matched to the template, i.e. produces large OHD. Transferring the points from that cube into the hole would not make sense, and it seems more reasonable to try matching with a smaller template size.

To account for the cases mentioned above, we extend template matching with adaptive cube size (ACS) selection. Let $\psi_\mathbf{q}^n$ be a cube of size $n \times n \times n$ voxels, centered at point $\mathbf{q}$, with edges parallel to the $x, y, z$ axes. The smallest cube size we consider is $n = 5$. The initial priority is calculated and the highest priority point $\widehat{\mathbf{p}}$ is found with $n = 5$. Following the procedure from Section II, we find the best match $\psi_{\widehat{\mathbf{q}}}^n \in \Phi$ for $\psi_{\widehat{\mathbf{p}}}^n$. Let the rotated and aligned $\psi_{\widehat{\mathbf{q}}}^n$ be denoted $\overline{\psi^n}_{\widehat{\mathbf{q}}}^{\mathbf{R}_b}$, and let the OHD between this cube and the template be $d\left(\overline{\psi^n}_{\widehat{\mathbf{q}}}^{\mathbf{R}_b}, \psi_{\widehat{\mathbf{p}}}^n\right) = \epsilon$. This $\epsilon$ gives us an idea of what kind of matching error we may hope to find for larger cubes. Based on this, we set a matching threshold $T = 1.0001 \cdot \epsilon$.

**Algorithm 1** Matching with Adaptive Cube Size
1: Find $\widehat{\mathbf{p}} = \arg\max_{\mathbf{p} \in \delta\Omega^t} P(\mathbf{p})$ and set $n = 5$
2: Find the best-matched cube $\psi_{\widehat{\mathbf{q}}}^n$ according to (2)-(4)
3: Let $\epsilon = d\left(\overline{\psi^n}_{\widehat{\mathbf{q}}}^{\mathbf{R}_b}, \psi_{\widehat{\mathbf{p}}}^n\right)$ and $T = 1.0001 \cdot \epsilon$
4: Initialize $\mathcal{C} = \left\{\psi_{\mathbf{q}_i}^n \in \Phi : d\left(\overline{\psi^n}_{\mathbf{q}_i}^{\mathbf{R}_b}, \psi_{\widehat{\mathbf{p}}}^n\right) \leq T\right\}$
5: **while** $|\mathcal{C}| > 1$ **do**
6: $\quad n \leftarrow n + 2$
7: $\quad$ **for** All $\psi_{\mathbf{q}_i}^n \in \mathcal{C}$ **do**
8: $\quad\quad$ **if** $d\left(\overline{\psi^n}_{\mathbf{q}_i}^{\mathbf{R}_b}, \psi_{\widehat{\mathbf{p}}}^n\right) > T$ **then**
9: $\quad\quad\quad$ Remove $\psi_{\mathbf{q}_i}^n$ from $\mathcal{C}$
10: $\quad\quad$ **end if**
11: $\quad$ **end for**
12: **end while**
13: **if** $|\mathcal{C}| == 1$ **then**
14: $\quad$ The cube in $\mathcal{C}$ (of size $n$) is the best match
15: **else**
16: $\quad$ Best-matched cube of the previous size is the best match
17: **end if**

After setting $T$, we create a set $\mathcal{C} = \{\psi_{\mathbf{q}_1}^n, \psi_{\mathbf{q}_2}^n, ..., \psi_{\mathbf{q}_N}^n\}$ of all cubes of size $n$ in $\Phi$, such that $d\left(\overline{\psi^n}_{\mathbf{q}_i}^{\mathbf{R}_b}, \psi_{\widehat{\mathbf{p}}}^n\right) \leq T$. This is a set of candidate cubes that match well with the template of size $n$. Then we increase $n$ and eliminate from $\mathcal{C}$ all cubes whose matching error exceeds $T$ with the new size. The process repeats until we are in a position to make a decision about the best-matched cube. Specifically, we keep increasing $n$ until $\mathcal{C}$ either ends up with only one cube (which is then the best match) or becomes empty. In the latter case, we go back to the previous (lower) cube size and select the cube with the lowest matching error as the best match. The procedure is summarized in Algorithm 1, where $|\mathcal{C}|$ represents the number of elements in $\mathcal{C}$. In our implementation, the cube size $n$ increases in steps of 2, as indicated in step 6 of the algorithm, but other increment schedules are possible.

*B. Non-Rigid Transformation*

Let $\overline{\psi^n}_{\widehat{\mathbf{q}}}^{\mathbf{R}_b}$ the best-matched cube found above, after rotation to align it with $\psi_{\widehat{\mathbf{p}}}^n$. As in Section II, let $M_{\widehat{\mathbf{q}}}$ be the set of points in $\overline{\psi^n}_{\widehat{\mathbf{q}}}^{\mathbf{R}_b}$ that are matched with points in $\psi_{\widehat{\mathbf{p}}}^n$, according to (5). Further improvement in the alignment of $\overline{\psi^n}_{\widehat{\mathbf{q}}}^{\mathbf{R}_b}$ and $\psi_{\widehat{\mathbf{p}}}^n$ can be achieved by Non-Rigid Transformation (NRT). The main challenge in developing a NRT in this case is finding the appropriate transformation for the unmatched points (i.e., those in $\overline{\psi^n}_{\widehat{\mathbf{q}}}^{\mathbf{R}_b} \setminus M_{\widehat{\mathbf{q}}}$). To overcome this challenge, we propose the following strategy: the matched points (those in $M_{\widehat{\mathbf{q}}}$) will be transformed to minimize their distance to their matches in $\psi_{\widehat{\mathbf{p}}}^n$, while the unmatched points will be transformed in a way that encourages the smoothness of transformation among neighboring points in $\overline{\psi^n}_{\widehat{\mathbf{q}}}^{\mathbf{R}_b}$.

The NRT we will apply to $\overline{\psi^n}_{\widehat{\mathbf{q}}}^{\mathbf{R}_b}$ is a collection of affine transformations $\{\mathbf{T}_i\}$, where $\mathbf{T}_i$ is a $3 \times 3$ matrix applied to $\mathbf{y}_i \in \overline{\psi^n}_{\widehat{\mathbf{q}}}^{\mathbf{R}_b}$. For the matched points, we define the *distortion* term

$$D = \sum_{\mathbf{y}_i \in M_{\widehat{\mathbf{q}}}} \|\mathbf{x}_i - \mathbf{T}_i \mathbf{y}_i\|_2^2, \qquad (6)$$

where $\mathbf{x}_i \in \psi_{\widehat{\mathbf{p}}}^n$ is the point in $\psi_{\widehat{\mathbf{p}}}^n$ matched with $\mathbf{y}_i$ according to (5). In order to take unmatched points into account, we construct an undirected $k$-nearest neighbor graph ($k = 5$ in our case) with points in $\overline{\psi^n}_{\widehat{\mathbf{q}}}^{\mathbf{R}_b}$ as vertices. Let $E$ be the set of edges in this graph. Then we define the *smoothness* term

$$S = \sum_{(\mathbf{y}_i, \mathbf{y}_j) \in E} \|\mathbf{T}_i - \mathbf{T}_j\|_F, \qquad (7)$$

where $\|\cdot\|_F$ is the Frobenius norm. Finally, the total cost is the combination of the distortion term and the smoothness term,

$$J = D + \lambda S, \qquad (8)$$

where $\lambda$ is the parameter that allows a trade-off between $D$ and $S$ in the cost function. In our experiments, we set $\lambda = 1$.

If we define a block matrix $\mathbf{T} = [\mathbf{T}_1 \ldots \mathbf{T}_{N_2}]^T$, where $N_2$ is the number of points in $\overline{\psi^n}_{\widehat{\mathbf{q}}}^{\mathbf{R}_b}$, then, following [17], $J$ is a quadratic function of $\mathbf{T}$. Hence, $\mathbf{T}$ that minimizes $J$ can be found in closed-form, and the complexity of computing it is $\mathcal{O}(N_2^3)$. For details, please refer to section 4.2 in [17]. After optimal $\mathbf{T}_i$'s are found, they are applied to all points in $\overline{\psi^n}_{\widehat{\mathbf{q}}}^{\mathbf{R}_b}$. Then the unmatched points of the transformed $\overline{\psi^n}_{\widehat{\mathbf{q}}}^{\mathbf{R}_b}$ are found and transferred to the hole, as described in Section II.

## IV. EXPERIMENTAL RESULTS

We test the proposed hole filling framework on 3D point clouds from two datasets: the Microsoft Voxelized Upper Bodies [18] and the Stanford 3D Scanning Repository [19]. There are mesh models in the Stanford dataset and here we stripped away the mesh connectivity and used vertices as a point cloud. However, the models in the Microsoft data set are real point clouds captured by four depth cameras. Both qualitative and quantitative results are presented and compared to those of [13], [9], [10], [2]. For this purpose, we implemented [9], [10], [2] following the papers.

Holes were generated in 3D point clouds by removing all points in a relatively large parallelepiped whose sides were aligned with $x$, $y$, and $z$ axes. Hole filling results for such generated holes are shown in Fig. 2 for two point cloud models from the Stanford dataset. In this figure, the point clouds are rendered using MeshLab software tool [20]. The first row shows the original point cloud/surface, second row shows the hole, while rows three to eight show the results of hole filling by the method in [13] (referred to as Base), [13] extended with ACS (Base+ACS), [13] extended with both ACS and NRT (Base+ACS+NRT) and those of [9], [10], [2], respectively. As seen in the figure, both Base+ACS and Base+ACS+NRT improve the reconstruction of the underlying surface compared to Base. All these three methods provide better reconstruction than [9], [10], [2]. The results in the first column (Armadillo) show that Base+ACS and Base+ACS+NRT provide fairly good known that exemplar-based methods are capable of generating plausible (but not necessarily correct) reconstruction of missing regions, through iterative transfer of small patches from elsewhere. Some theoretical support for such findings may be found in the recently proposed low dimensional manifold model (LDMM) [21], although we do not extend that model to point clouds here.

TABLE I
THE AVERAGE NSHD ± STANDARD DEVIATION ($\times 10^{-7}$)

| Point cloud | Base | Base+ACS | Base+NRT | Base+ACS+NRT | [9] | [10] | [2] |
|---|---|---|---|---|---|---|---|
| Andrew | 0.91 ± 0.17 | 0.87 ± 0.13 | 0.85 ± 0.14 | **0.81 ± 0.11** | 2.19 ± 0.31 | 2.12 ± 0.31 | 20.14 ± 3.77 |
| Ricardo | 0.45 ± 0.13 | 0.33 ± 0.11 | 0.43 ± 0.13 | **0.31 ± 0.12** | 0.92 ± 0.28 | 0.91 ± 0.24 | 4.78 ± 0.33 |
| David | 0.78 ± 0.14 | 0.73 ± 0.13 | 0.71 ± 0.14 | **0.69 ± 0.11** | 2.34 ± 0.30 | 2.09 ± 0.32 | 18.24 ± 2.67 |
| Phil | 0.45 ± 0.13 | 0.32 ± 0.10 | 0.43 ± 0.13 | **0.31 ± 0.12** | 0.94 ± 0.33 | 0.83 ± 0.21 | 5.13 ± 0.42 |
| Bunny | 4.23 ± 0.89 | 2.55 ± 0.47 | 4.01 ± 0.63 | **2.21 ± 0.39** | 7.89 ± 1.12 | 7.34 ± 1.03 | 20.71 ± 2.93 |
| Happy Buddha | 2.87 ± 0.42 | 1.36 ± 0.41 | 2.62 ± 0.38 | **1.13 ± 0.27** | 6.33 ± 0.92 | 6.94 ± 1.02 | 24.77 ± 3.71 |
| Armadillo | 4.37 ± 1.12 | 2.68 ± 0.39 | 3.89 ± 0.97 | **2.04 ± 0.21** | 9.91 ± 1.82 | 9.13 ± 1.81 | 21.68 ± 3.36 |
| Dragon | 3.15 ± 0.85 | 1.86 ± 0.38 | 2.81 ± 0.65 | **1.47 ± 0.33** | 11.82 ± 1.94 | 7.94 ± 1.51 | 19.42 ± 3.58 |
| Thai Statue | 6.38 ± 0.96 | 4.09 ± 0.73 | 6.08 ± 0.91 | **3.78 ± 0.69** | 18.24 ± 3.31 | 19.09 ± 4.71 | 38.13 ± 6.41 |
| Lucy | 7.28 ± 0.91 | 4.16 ± 0.75 | 7.08 ± 0.91 | **2.73 ± 0.56** | 19.17 ± 3.82 | 21.13 ± 4.71 | 33.12 ± 6.94 |
| Asian Dragon | 5.38 ± 0.96 | 4.14 ± 0.72 | 5.11 ± 0.91 | **3.12 ± 0.61** | 18.24 ± 3.29 | 19.13 ± 4.76 | 30.13 ± 6.14 |

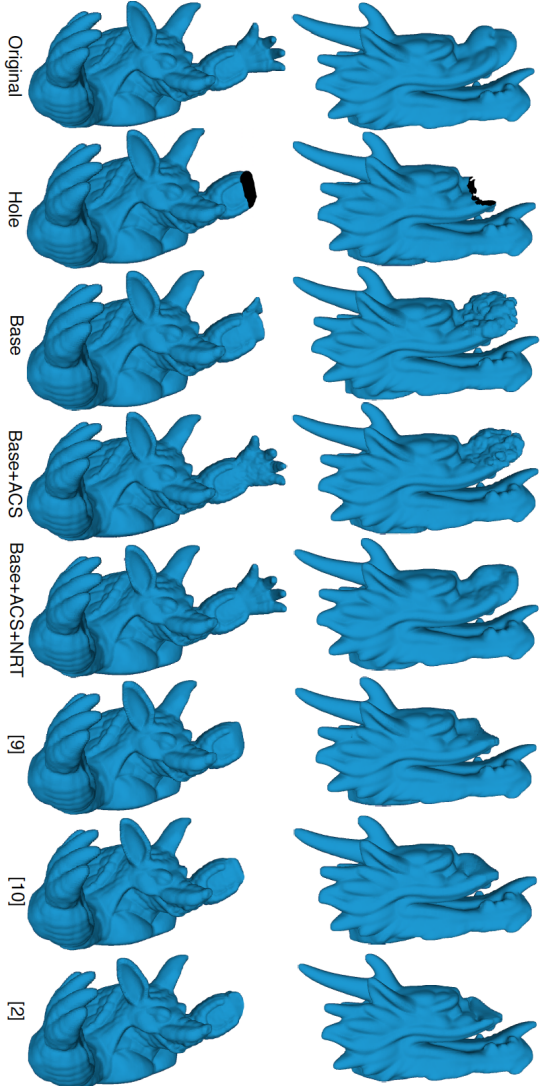

Fig. 2. Hole filling results on Armadillo (column 1) and Dragon (column 2)

We also show quantitative comparison using the Normalized Symmetric Hausdorff Distance (NSHD) between the reconstructed set of points (denoted $S_r$) and the original set of points (denoted $S_o$):

$$d_s(S_r, S_o) = \frac{1}{V} \max\{d(S_r, S_o), d(S_o, S_r)\}, \qquad (9)$$

where $V$ is the volume of the smallest axes-aligned parallelepiped enclosing the given 3D point cloud. Tests were performed on four point cloud models from the Microsoft dataset (first four models in Table I) and seven models from the Stanford dataset. In each 3D point cloud, we randomly selected the locations of 15 holes (punched by a parallelepiped whose dimensions were, on average, 20% of the range of data in each direction) and filled them using the various hole-filling methods. Table I shows the average NSHD (± standard deviation) of the 15 test cases. In addition to the six methods from Fig. 2, we also show quantitative results for Base+NRT, which is [13] extended with NRT, but without ACS.

As seen in the results, both ACS and NRT are able to bring improvements to Base. Specifically, Base+NRT is slightly more accurate than Base, while Base+ACS is considerably more accurate than Base. But the best accuracy (indicated in bold) is achieved when ACS and NRT are combined (Base+ACS+NRT). All these methods offer more accurate reconstruction compared to those of [9], [10], [2]. Runtime results are presented in the supplement. Based on those results, [2] is the fastest of the seven methods, followed by Base+ACS, Base+ACS+NRT, Base, Base+NRT, [9], and finally [10]. The results also show that NRT takes up to 8% of the total run time in Base+NRT and Base+ACS+NRT.

reconstruction of the left hand, benefiting in part from the symmetry that exists with the right hand in the model. The second column (Dragon) is more challenging, since the nose does not have a symmetric counterpart in this model. Nonetheless, Base+ACS+NRT provides a fairly plausible reconstruction (Base and Base+ACS are somewhat less plausible), while [9], [10], [2] simply cut the nose off. A zoomed-in results of the hole region in Fig. 2 and other additional examples are provided in the supplementary document. These findings are consistent with experience in image inpainting [14], where it is

## V. CONCLUSION

We have presented two improvements to the exemplar-based framework for 3D point-cloud hole filling: adaptive cube size selection and non-rigid transformation. The new framework offers better accuracy and competitive execution time compared to other recent proposals for 3D point-cloud hole filling. In the future, we plan to extend this framework for color inpainting of 3D point clouds.